\documentclass[doublecol]{epl2}

\usepackage{graphicx}
\usepackage{epsfig}

\usepackage{amsmath}
\usepackage{amsthm}
\usepackage{amssymb}

\usepackage{bm}
\usepackage{hyperref}

\usepackage[english]{babel}

\usepackage[latin1]{inputenc}
\usepackage{dcolumn}

\title{Temperature $\times$ Doping Phase Diagram of Cuprate Superconductors}

\author{Lizardo H. C. M. Nunes\inst{1}, A. W. Teixeira \and E. C. Marino\inst{2} }
\shortauthor{Nunes \etal}
\institute{
  \inst{1} Departamento de Ci\^encias Naturais, Universidade Federal de S\~ao Jo\~ao del Rei, 36301-000 S\~ao Jo\~ao del Rei, MG, Brazil\\
  \inst{2} Instituto de F\'{\i}sica, Universidade Federal do Rio de Janeiro, Caixa Postal 68528, Rio de Janeiro, RJ, 21941-972, Brazil
}


\abstract{
Starting from a spin-fermion model for the cuprate superconductors,
we obtain an effective interaction for the charge carriers
by integrating out the spin degrees of freedom.
Our model predicts a quantum critical point for the superconducting interaction coupling,
which sets up a threshold for the onset of superconductivity in the system.
We show that the physical value of this coupling is below this threshold, thus
explaining why there is no superconducting phase for the undoped system.
Then, by including doping, we find a dome-shaped dependence of the critical temperature
as charge carriers are added to the system,
in agreement with the experimental phase diagram.
The superconducting critical temperature
is calculated without adjusting any free parameter and yields,
at optimal doping  $ T_c \sim 45 $ K, which
is comparable to the experimental data. }

\begin{document}

\maketitle

\section{Introduction}\label{Introduction}

So far, there is no consensus concerning the microscopic mechanism which is responsible for the appearance of superconductivity in the cuprate superconductors.
Yet, it is widely accepted that some sort of spin exchange mechanism should be responsible for the Cooper pair formation in the cuprates \cite{scalapino2012,pines,moriya2006,muramatsu1988}. Following this path, in the present paper we obtain a novel superconducting interaction for the cuprates, which provides critical temperatures
(high-$ T_c $ ) that are comparable to the ones experimentally observed,
without resorting to any  adjustment of free parameters.

We also obtain a dome shaped superconducting phase diagram as charge carriers are added to the system, which reproduces qualitatively the experimental behaviour observed for the cuprates as the compound is doped. Indeed, the parent compounds of high-$ T_c $ superconductors are Mott insulators and the system becomes superconducting as holes are pumped into the
CuO$_2$ planes while extra atoms (oxygen in YBCO, strontium in LSCO) are stoichiometrically added to the system. $ T_ c $ increases up to an optimal value and than decreases forming the characteristic dome shaped phase diagram \cite{wen2006,dagotto1994}.

Our starting point is the spin-fermion model, which has been extensively used to describe the cuprates previously \cite{efetov2013,berg2010,abanov,kotliar1990}. The task of providing a minimal Hamiltonian, with only a few parameters, which captures the main physics presented by the cuprates can be rather elusive, because key issues may be lost in the attempt of simplifying the system description. As a matter of fact, there is no consensus whatsoever regarding the minimal model which entails the vast phenomenology presented by the cuprates, however the three-band Hubbard model proposed by Emery  \cite{emery1,emery2} is a good candidate to model the CuO$_2 $ planes of the cuprates. It is well established that it is from these planes that superconductivity emerges in the cuprates. However, the three-band model model is given in terms of several parameters and its analysis can be rather complicated. On the other hand, in the absence of doping the parent compound is reduced to a single band Hubbard model with a strong on-site Coulomb repulsion at half filling. For the square lattice, which is the appropriate lattice topology for the CuO$_2 $ planes, the model is mapped into the antiferromagnetic spin 1/2 Heisenberg model \cite{fazekas},
which, indeed, describes well the dynamics of the spin degrees of freedom of undoped copper oxides \cite{Manousakis1991}.
Moreover,  analytic calculations for the Heisenberg model provided the basis for understanding a range of experimental results for the undoped cuprates \cite{Chakravarty1988,Arovas1988}. As the system is doped, charge carriers are added to the $ p $ orbitals of the oxygen and we assume that the localized spins of the Cu should interact with the spins of the itinerant fermionic charge carriers. This is the picture that we envision for the calculation of the superconducting phase diagram in the present paper.
In fact, several authors have employed the same approach in order to model the cuprates superconductors previously~\cite{Cancrini1991}-\cite{Catellani1988}.

Our approach is different from those previous attempts to describe the phenomenology of the cuprates because presently we employ the spin coherent states to integrate out the spin degrees of freedom in order to obtain the effective interaction for the charge carriers. It should be emphasized that we did not resort to any kind of perturbative method, diagrammatic expansion or auxiliary slave-boson technique in order to obtain our effective fermion theory.
Next,
we calculate the superconducting phase diagram
using as input the physical values of parameters, which have been measured for the cuprates.

The paper is divided as follows:
in the next section we introduce the model
and briefly outline the derivation of the effective dynamics for the itinerant fermion fields.
Then, we proceed by calculating the $ T_c \times \mu$, superconducting phase diagram
obtaining the familiar dome shaped  diagram,
which displays values that are comparable to the ones experimentally observed.
However, for systems in a spatial dimension less than three
the Coleman-Mermin-Wagner-Hohenberg theorem \cite{mw}
forbids the occurrence of a phase transition at finite temperatures.
Nonetheless, for two dimensional systems,
such as our case,
there is an underlying Berezinskii-Kosterlitz-Thouless (BKT) transition \cite{kt}
for $ T < T_{KT} $,
below which phase coherence is found for a nonzero order parameter.
The actual temperature for the appearance of superconductivity
is set at $ T_{ KT } $, where $ T_{ KT} \leq T_c $ \cite{babaev}.
As should be expected, for a very large number of superconducting planes,
$ T_{KT} \rightarrow T_c $.
Therefore the superconducting critical temperature calculated in the present paper
may be regarded as a mean-field critical temperature for the KT transition.
In Sec. \ref{Conclusions} our conclusions are presented.
In the appendix we discuss how $ d $-wave superconductivity can be obtained

\section{The effective model}\label{TheModel}

Consider a single CuO$_2$ plane containing both localized spins, wich are located at the sites of a square
lattice and charge carriers with a tight-binding dispersion relation, assumed to be Dirac-Like
\cite{Wehling2014}.
The localized spins interact with the spin degrees of freedom of the charge carriers by a Kondo-like term,
whereas the localized spins mutual interaction is described by an antiferromagnetic Heisenberg Hamiltonian.
The complete Hamiltonian is
\begin{eqnarray}
H
& =  &
- t \sum_{ <ij> }
\left(
c^{ \dagger }_{ i \alpha } c_{ j\alpha}+hc
\right)
\nonumber \\
& &
+J_{K} \sum_{ i } {\bf S}_i \cdot
\left(
c^{ \dagger }_{ i \alpha }\vec{\sigma}_{\alpha\beta} c_{ j\alpha}
\right)
\nonumber \\
& &
+J_{d} \sum_{ <ij> } {\bf S}_i \cdot {\bf S}_j
\,  ,
\label{EqHSpin-Fermion}
\end{eqnarray}
where $ {\bf S}_i $ is the localized spin operator
and $ c^{\dagger}_{i\alpha} $ is the creation operator of a charge carrier
with spin $ \alpha = \uparrow, \downarrow $,
both at site $ i $.
In terms of the three-band Hubbard model parameters,
the Kondo coupling of an itinerant oxygen hole spin
and the nearest local Cu spin is
\cite{Zannem1988,Kampf1994}
\begin{equation}
J_{k}
=
t_{pd}^{2}
\left(
\frac{1}{\Delta E} + \frac{1}{U_{d}+\Delta E}
\right)
\end{equation}
and the exchange coupling between the Cu magnetic moments is given by
\cite{Zannem1988,Kampf1994}
\begin{equation}
J_{d}
=
\frac{4 t_{pd}^{4}
}{
\left(\Delta E + U_{pd}\right)^{2}
}
\left(
\frac{1}{ U_{d}} + \frac{1}{2 \Delta E + U_{p} }
\right) \, .
\label{Jd}
\end{equation}
Presently, in numerical calculation we employ the following values for the above parameters
\cite{coupling1}:
$ t_{pd}= 1.48 $, $ U_{d} = 8.5 $, $ U_p = 4.1 $, $ U_{pd} = 1.3 $
and the difference in energies \cite{coupling2} are $ \Delta E = U_{d}/4 - U_{p}/8 = 1.61 $,
all given in units of eV.

Now, we express the partition function
as a path integral in the complex time representation.
In order to obtain the continuum limit of our model Hamiltonian in (\ref{EqHSpin-Fermion}),
we use spin coherent states. In the presence of these, we may
replace the spin operators $ {\bf S}_i $ by
$ S{\bf N}({\bf x}) $,
where $ S $ is the spin quantum number and
$ {\bf N}({\bf x}) $ is a classical vector that is decomposed into two perpendicular components
associated with ferromagnetic and antiferromagnetic fluctuations respectively,
$ {\bf L} $ and $ {\bf n} $.
Moreover, we also replace the operator $ c^{\dagger}_{ i \alpha} $
by the fermion field $ \psi^{ \dagger }_{ \sigma } ( x ) $,
as usual.
Hence, the partition function becomes,
\begin{eqnarray}
\mathcal{ Z }
& = &
\int
\mathcal{ D } \psi  \mathcal{ D } \psi ^{\dagger}
\mathcal{D} {\bf L} \mathcal{D} {\bf n} \,
\delta
\left[
|{\bf n}|^{2}-1
\right]
\nonumber \\
& &
\exp
 \left[
-\int_{0}^\beta
d\tau \int
d^{2}x
 \left(
\mathcal{H} - \psi^{\dagger} i\partial_{\tau} \psi
\right) \right]
\, ,
\label{EqPartition}
\end{eqnarray}
where the continuum Hamiltonian density reads
\begin{eqnarray}
\mathcal{H}
& = &
\psi^{\dagger}
\left(
i \,  \hbar v_{f}  \vec{\sigma}\cdot \vec{\nabla} - \mu
\right)
\psi + \frac{ \rho_{s} }{ 2 } |a \nabla {\bf n}|^{2}
\nonumber \\
& &
+ \frac{ \chi_{\bot} }{ 2 } S^{2}
|{\bf L}|^{2}
+S {\bf L} \cdot
\left[
J_{K} {\bf s} + i
\left(
{\bf n} \times \partial_{\tau} {\bf n}
\right)
\right]
\nonumber \\
& &
+(-1)^{|x|}SJ_{K}{\bf n}\cdot{\bf s}
\, ,
\label{EqH1}
\end{eqnarray}
with the itinerant spin operator as
$ {\bf s} = \psi_{\delta}^{\dagger}(\vec{\sigma})_{\delta \gamma}\psi_{\gamma}$,
written in terms of the Pauli matrices,
$ \vec{\sigma} = ( \sigma_x , \sigma_y , \sigma_z ) $,
and the fermion field has spinorial components
$ \psi^{\dagger }_{\alpha} = ( \psi^{\dagger }_{1 \alpha}, \psi^{\dagger }_{2 \alpha} ) $.
The spin stiffness is given by $ \rho_{s}=J_{d}S^{2}$,
and the transverse susceptibility is $ \chi_{\bot} = 4 J_{d} $.
Notice that the chemical potential $ \mu $ controls the total number of charge carriers
that are added to the conduction band as the system is doped.
Moreover, the parameters values for the YBCO are known
and given by \cite{Marino2001}
$ {\hbar}v_{f}=1.15~eV\AA $ and
$ a = 2.68\sqrt{2}~\AA $,
which are employed in the present paper for the numerical analysis.

We can perform the gaussian integration over $ {\bf L} $ in (\ref{EqPartition})
and rewrite the AF fluctuation field $ n_{i} $
using the $ CP^{1} $ formulation (Schwinger bosons) of the O(3) nonlinear sigma model,
which is written in terms of the two complex fields $ z_{\alpha} $ ($\alpha=1,2 $),
namely
$ n_{ i } =  z^{ * }_{ \alpha } \left( \sigma_{ i } \right)_{ \alpha \beta } z_{ \beta } $,
$ i = x,y, z $ \cite{auerbach}.
We also perform the canonical transformation on the fermion field \cite{Muramatsu1993}
$
\psi_{\alpha} \rightarrow U_{\alpha \beta} \psi_{\beta}
$,
where the unitary matrix $ U $ is given in terms of the $ z_{\alpha} $-fields,
\begin{equation}
U =
\left(
 \begin{array}{cccc}
 z_{ 1 } &  -z^{*}_{ 2 } \\
 z_{ 2 } &  -z^{*}_{ 1 }
\end{array}
\right) \,  .
\label{EqU}
\end{equation}
Employing the polar representations of the bosonic fields,
$ z_{ \alpha } = \rho_{ \alpha } e^{ i \, \theta_{ \alpha }  } / \sqrt{ 2 }  $,
we can perform the functional integration over the Schwinger boson fields
assuming constant $ \rho_{ \alpha } $
and we obtain the resulting effective Lagrangian density for the fermion fields associated to the charge carriers \cite{AOP},
\begin{eqnarray}
\mathcal{ L }_{  \mbox{\scriptsize{eff}} }
& = &
\psi^{ \dagger }  \left[ i\gamma^0\gamma^\mu\partial_\mu -\mu\right] \psi
+
g_{0} \, p^{ \dagger } p
\nonumber \\
& &
+ g_{1} \, \left( \bar\psi_\sigma \psi_\sigma \right)^2
+ g_{2}\, \left( \bar{\psi}_{\sigma} \gamma^{0} \psi_{\sigma} \right)^{2}
\nonumber \\
& &
+ g_{3} s_{z}^{2}
\, ,
\label{Eq_Leff}
\end{eqnarray}
where the above interaction strengths are
$ g_{0}=(2 \tilde{\rho}_{s})^{-1} $,
$ g_{1} = (8\tilde{\rho}_{s})^{-1} $,
$ g_{2} =
\left(
4 J_{k}^{2}\tilde{\rho_{s}}/ \chi_{\bot} - 1
\right)
( 8\tilde{\rho}_{s})^{-1} $,
and
$
g_{3} =
\left(
 c^{2}/\rho_{s}- 4J_{K}^{2}/ \chi_{\bot}
\right) /8 $, with $ c^{2} = \rho_{s} \chi_{\bot} $ and $ \tilde{\rho}_{s} = \rho_{s} (a/\hbar v_{f})^{2} $.
The first interaction term corresponds to a BCS-type superconducting interaction,
with the singlet pair operator denoted by
$
p = \psi_{2\downarrow } \ \psi_{1\uparrow } + \psi_{1\downarrow } \ \psi_{2\uparrow }
$,
the second and third terms produce an insulating charge-gapped phase showing an excitonic condensate,
a Nambu-Jona-Lasinio-type,
the fourth term correspond to an anisotropic spin-spin interaction.
The couplings above are calculated taking the parameters from our model Hamiltonian,
which yields,
$ g_{0}^{ \mbox{\scriptsize{(est)}} } = 0.21  $ eV,
$ g_{1}^{ \mbox{\scriptsize{(est)}} } = 0.052 $ eV,
$ g_{2}^{ \mbox{\scriptsize{(est)}} } = 0.34 $ eV,
and
$ g_{3}^{ \mbox{\scriptsize{(est)}} } = 0.045 $ eV,
and we see that $ g_1^{ \mbox{\scriptsize{(est)}} }  $
and $
g_3^{ \mbox{\scriptsize{(est)}} }  $
are small compared to the superconducting interaction strength.

One can now introduce the Hubbard-Stratonovich (HS) auxiliary fields in (\ref{Eq_Leff}),
\begin{equation}
\Delta =
g_0 \,
\left(
\psi_{2\downarrow}\psi_{1\uparrow}+\psi_{1\downarrow}\psi_{2\uparrow}
\right) \, ,
\label{Eq_Delta}
\end{equation}
\begin{equation}
M
= g_1 \,
\bar{\psi}_{\sigma}\psi_{\sigma} \, ,
\label{Eq_M}
\end{equation}
\begin{equation}
N
= g_2 \,
{\psi}^{ \dagger }_{\sigma}\psi_{\sigma} \, ,
\label{Eq_N}
\end{equation}
\begin{equation}
s_{z}
= g_3 \,
\psi_{\gamma}^{\dagger}\sigma_{\gamma\delta}^{z}\psi_{\delta} \, ,
\label{Eq_sz}
\end{equation}
in order to rewrite
the effective Lagrangian density
as
\begin{equation}
\mathcal{L}_{ \mbox{\scriptsize{eff, HS}} }
=
-
\int
\frac{d^{2}k}{
\left(
2\pi
\right)^{2}
}
\; \Phi^{\dagger}(k) \, \mathcal{A} \, \Phi(k) \, ,
\label{Eq_LeffHS}
\end{equation}
where the Nambu field is given by
\begin{equation}
\Phi^{\dagger}(k) =
\left( \,
\psi^{\dagger}_{1,\uparrow}(k) \,
\psi^{\dagger}_{2,\uparrow}(k) \,
\psi_{1,\downarrow}(-k) \,
\psi_{2,\downarrow}(-k) \,
\, \right)
\label{Eq_Nambu}
\end{equation}
and the matrix $ \mathcal{ A } $ is
\begin{equation}
\left(
 \begin{array}{cccc}
 - \mu_{ + } + N_{ - } & -\hbar v_{F}k_{-} & 0 & -\Delta \\
 -\hbar v_{F} k_{+} &  -\mu_{ - } + N_{ - } & -\Delta & 0 \\
 0 & -\Delta^{*} & \mu_{ + } - N_{ + } & \hbar v_{F}k_{+} \\
 -\Delta^{*} & 0 & \hbar v_{F}k_{-} & \mu_{ - } - N_{ + }
\end{array}
\right) \, ,
\label{Eq_A}
\end{equation}
such that $ k_{\pm}=k_{y}\pm i k_{x} $,
$ \mu_{\pm} = \mu {\pm} M $
and $ N_{\pm} = N {\pm} s_z $,
since we have Fourier transformed  $ \mathcal{L}_{ \mbox{\scriptsize{eff, HS}} } $,
and the standard quadratic terms of the HS fields have been omitted in (\ref{Eq_LeffHS}) for the sake of simplicity.
This transformation is exact and no approximation has been performed to the theory up to this point.
Notice that we have eliminated the quartic fermionic interactions in (\ref{Eq_LeffHS})
at the expenses of having introduced the scalar fields (\ref{Eq_Delta})-(\ref{Eq_sz}).
But now we can perform the Gaussian integration over the fermionic fields exactly,
since the partition function can be expressed as a path integral in the complex time representation.

Replacing the HS fields for their expected values,
the auxiliary field $ N $ in (\ref{Eq_N}) only introduces a trivial shift of the chemical potential
and therefore shall be omitted from now on.
The case for $ g_0, g_3 \neq 0 $ has been previously reported \cite{heron}.
However, in the present paper, we shall neglect the $ g_3 $ interaction term,
since its coupling is small compared to $ g_0 $
and we are not interested in the magnetic ordering of the itinerant fermion fields for the cuprates,
only the superconducting phase is considered in the present paper.
Furthermore,
to our best knowledge, the case $ g_0 , g_1, g_3 \neq 0 $
has never been investigated, but the case for $ g_3 = 0 $ has been previously reported \cite{PLA}
and it was shown that, as long as $ g_0 > g_1 $,
such as the case for our parameter values,
the system does not present the excitonic gap and becomes superconducting
as charge carrier are added to the system,
$ \mu > 0 $ increases.

\section{The superconducting phase diagram}\label{SCPhase}

We start our analysis calculating the free energy (effective potential) as a function of the order parameters.
We proceed to the evaluation of the partition function as a path integral in the complex time representation and integrate it over the fermionic fields.
Thereby, the effective potential becomes
\begin{eqnarray}
V_{ \mbox{\scriptsize{eff}} }
& = &
-\frac{1}{\beta} \sum_{n=-\infty}^{\infty}\left(\frac{a_{D}}{2 \pi}\right)^{2}
\nonumber \\
&&
\int d^{2}k
\log
\left(
\frac{\prod_{j=1}^{4}[i\omega_{n}-E_{j}]}
{\prod_{j=1}^{4}[i\omega_{n}-E_{j}(M=0,\Delta=0)]}
\right)
\nonumber \\
&  &
+ \frac{|\Delta |^{2}}{ g_0 }
+
\frac{M^{2}}{ g_1 }
\,  ,
\label{EqEffPotencial}%
\end{eqnarray}
where $\omega_{n} = (2n + 1)\pi T  $ are the Matsubara frequencies for fermions, $ a_{D} $ is the lattice spacing for dopants, and
\begin{eqnarray}
E_{j}
& = &
\pm
\left( M^2 + |\Delta |^2 + \mu ^2 +\left(\hbar v_{f} | k |\right)^2
\right.
\nonumber \\
&  &
\hspace{-0.5cm}
\pm
\left.
2 \sqrt{ M^2 (\, |\Delta |^2 + \mu ^2 \,) + \mu ^2 \left(\hbar v_{f} | k | \right)^2 }
\right)^{1/2}
.
\label{EqEj}
\end{eqnarray}
The above expression corresponds to the first term of a large $ N $ expansion.
Any improvements to the critical temperature calculated here
can be done employing standard diagrammatic techniques.
However, our results are exact in the $ N \rightarrow \infty $ limit.

We start our analysis showing that at zero temperature and $ \mu = 0 $
there is a critical coupling $ g_c $
for the superconducting interaction strength
which is required for the system to become superconducting.
Indeed,
we integrate (\ref{EqEffPotencial}) over
$ k $ in the first Brillouin zone at $ T = 0 $
in order to obtain the effective potential in terms of $ \Delta $, with $ M = \mu = 0 $,
\begin{eqnarray}
V_{ \mbox{\scriptsize{eff}} } ( \Delta )
& = &
\frac{|\Delta |^{2}}{ g_0 }
+
\frac{a_{D}^{2}}{3 \pi (\hbar v_{f})^{2}}
\,
\left[
\,
\left(
|\Delta |^{2}+\left(\frac{\hbar v_{f} \pi}{a_{D}}\right)^{2}
\right)^\frac{3}{2}
\right.
\nonumber \\
& &
\left.
- |\Delta |^{3}-\left(\frac{\hbar v_{f} \pi}{a_{D}}\right)^{3}
\,
\right]
\,  .
\label{EqEffPotencial0}
\end{eqnarray}
The minima condition requires $ V_{ \mbox{\scriptsize{eff}} }' ( \Delta ) = 0 $
and $ V_{ \mbox{\scriptsize{eff}} }'' ( \Delta ) > 0 $.
Thus, $ \Delta_{0}=0 $ is a minimum only for $ g_{0} \le g_{c} $,
while, on the other hand,
\begin{equation}
\Delta_{0}
=
\frac{\left(
g_{0}a_{D}/(\hbar v_{f})
\right)^{2}-4}{4 g_{0} a_{D}^{2}/ \pi (\hbar v_{f})^{2}}
\label{EqDMinimun}
\end{equation}
is a minimum only for $ g_{0} > g_{c} $, with $ g_{c}=2 \hbar v_{f}/a_{D}=0.86 $ eV.
Comparing $ g_c $ and $ g_{0}^{ \mbox{\scriptsize{(est)}} } $,
we see that $ g_{0}^{ \mbox{\scriptsize{(est)}} } < g_c $,
explaining why the system is not superconducting
in the absence of doping (or $ \mu = 0 $),
as indeed observed experimentally.

We now turn to the finite temperature analysis.
Since we are looking for the condition of minima for the free energy,
we take the derivative of $V_{ \mbox{\scriptsize{eff}} } $ with respect to $ \Delta $ and
$ M $ after the summation over the Matsubara frequencies,
the nonzero solutions of the order parameters provide the following two coupled equations,
\begin{eqnarray}
\frac{1}{g_0 }=\sum_{j=\pm1} \left(\frac{a_{D}}{2 \pi}\right)^{2} \int d^{2}k
\frac{1}{2 \Delta}\frac{\partial \epsilon_{j}}{\partial \Delta}
\tanh \left( \frac{\beta}{2}\epsilon_{j}\right) \, ,
\end{eqnarray}
\begin{eqnarray}
\frac{1}{g_1 }=\sum_{j=\pm1} \left(\frac{a_{D}}{2 \pi}\right)^{2} \int d^{2}k
\frac{1}{2 M}\frac{\partial \epsilon_{j}}{\partial M}
\tanh \left( \frac{\beta}{2}\epsilon_{j}\right) \, .
\end{eqnarray}

\begin{figure}
[ht]
\centerline
{
\includegraphics[
angle=-90,
width=1.\columnwidth]{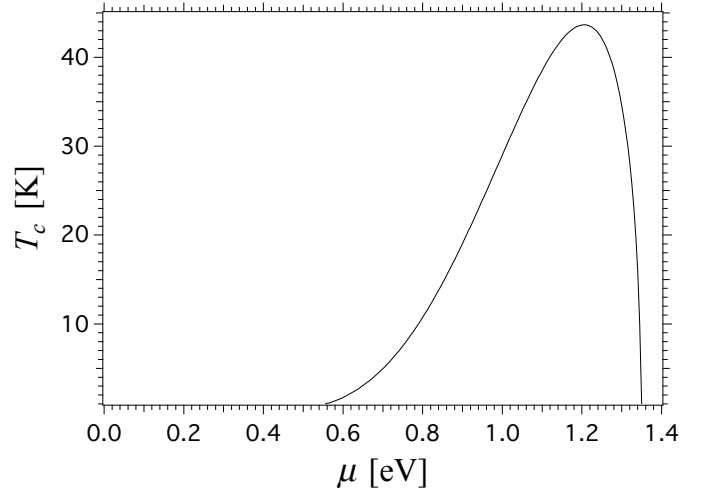}
}
\caption{$ T_c $ as a function of $ \mu $. }
\label{TcXMu}
\end{figure}

The numerical solutions for the superconducting critical temperature $ T_ c $
as a function of the chemical potential $ \mu $ can be seen in Fig.~\ref{TcXMu}.
Since $ g_0 > g_1 $,
we find only superconductivity for the system,
as we have already stated above,
and at $ \mu = 0 $ the system is in the normal state.
As $ \mu $ increases
a Fermi surface builds up and the system asymptotically becomes superconducting,
in agreement with the Cooper's theorem;
as $ \mu $ increases even further,
$ T_c $ reaches a maximum value at an optimal chemical potential
and it decreases as charge carriers are added to the system.
 A dome-shaped plot is consistent to previous results
for two-color and two-flavor QCD \cite{Fukushima2007}
and also strongly interacting two-dimensional Dirac fermions \cite{Smith2009}.
Those results and the phase diagram presently calculated
suggest that Dirac fermions may play a relevant role in the description
of systems containing Dirac fermions.
Moreover, we see that we have obtained a high value for $ T_c \sim 45 $ K,
which is comparable to the experimental data for a single CuO$_ 2 $ plane.
Notice that the high value of the critical temperature
was obtained simply employing the model parameters given above,
without resorting to any adjustment of free parameters.

\section{Conclusions}\label{Conclusions}

Starting from the spin-fermion model
we have obtained an effective model for charge carriers
by integrating out the spin degrees of freedom.

Our model predicts a critical interaction for the superconducting interaction
which sets a quantum critical point for the appearance of superconductivity in the system.
We have shown that the interaction strength is smaller than this threshold
explaining why there is no superconductivity in the absence of doping.
As charge carrier are added to the system,
a Fermi surface builds up and the quantum phase transition
is  washed out, in agreement with Cooper's theorem.

For the model parameters appropriate for the cuprates compounds,
we have calculated the superconducting phase diagram,
by means of the BKT mechanism,
and we have found a dome-shaped dependence of the temperature on the chemical potential,
which is in agreement with the well-known results for the cuprates.
Without resorting to any adjustment of free parameters,
we have found an optimal $ T_c $
which is high and comparable to the experimental data,
$ T_c \sim 45 $ K.
Superconductivity arises from a novel mechanism, which originates from
the purely magnetic interactions involving the localized spins and itinerant electrons
of the original system.
This result is also consistent with DMRG calculations for the 1d Heisenberg-Kondo model \cite{Xavier2008},
which shows the development of a superconducting phase mediated by antiferromagnetic fluctuations.
On the same token, mean-field calculations for the 2d Kondo lattice \cite{Liu2012},
shows that, as an AFM Heisenberg exchange coupling is taken into account,
singlet Cooper pair occurs among conduction electrons, leading to heavy fermion superconductivity.
Hence, our derivation may provide an analytical explanation for those results
without the resort of any approximation, which is a very interesting result.

The singlet pairing that we have obtained here
emerges from the interaction of local magnetic moments
with the spins of a conduction band.
Several compounds like
pnictides, heavy fermion compounds, chalcogenides (and cuprates as well)
may be described by multi-orbitals models
with interactions between magnetic moments and the spins of conduction bands.
Their structure, phase diagrams and experimental data
provide a phenomenological evidence relating these compounds \cite{Scalapino2012}.
Therefore, the approach employed here is not just restrained to the cuprates,
but might be applied to several other compounds,
providing a framework for the underlying microscopic mechanism
which is responsible for the appearance of superconductivity in several unconventional superconductors.

\acknowledgments
E. C. Marino has been supported in part by CNPq and FAPERJ.

\section{Appendix}\label{Appendix}

In this appendix we discuss how $ d $-wave superconductivity
can be obtained from the approach employed above.
We start pointing out that we have assumed
a constant contact point exchange coupling among local magnetic moments,
as seen in (\ref{Jd}).
Instead, let us take, for instance, the following exchange interaction \cite{abanov},
\begin{equation}
H_{ \rm exc }
=
g
\sum_{ \bf q } \chi^{-1}_0 ({\bf q})
\, S_{\bf q} S_{ -{\bf q} }
\, ,
\end{equation}
where $ g $ is some constant value and the spin susceptibility
was proposed by Millis, Monien and Pines \cite{Millis1988},
which has been shown to provide a quantitative fit to the NMR experiments in YBCO,
\begin{equation}
 \chi_0 ({\bf q}, \omega )
 =
 \frac{ \chi_{ 0 } }
 { 1 + \xi^2 \left( {\bf q} - {\bf Q } \right)^2  - i \omega / \omega_{ \rm SF }
 }
\, ,
\end{equation}
with $ \chi_{ 0 } $ as the static susceptibility peaked at wave vector
$ {\bf Q} = \left( \pi /a, \, \pi/a \right) $,
$ \xi $ as the antiferromagnetic correlation length
and $ \omega_{\rm SF } $ as the paramagnon energy.
This spin susceptibility have a regular Ornstein-Zernike form
and RPA calculations for nearly half filled Hubbard model
also provide the same overall form \cite{Scalapino2012}.

Following our approach, we argue that the superconducting coupling
becomes $ g_0 \rightarrow g_0 \chi( {\bf q } )$ in momentum space
and the new gap equation in the singlet channel becomes \cite{moriya2006}
\begin{equation}
\Delta( {\bf k } )
=
-
\bar{ g }
\int d^2 k
\,
\chi ( {\bf k } - { \bf p } )
\,
\Delta( {\bf p } )
\frac{ \tanh( \beta E / 2  ) }{ 2 E }
\, ,
\end{equation}
where $ \bar{ g } > 0  $.
Except for the extra minus sign in the r.h.s of the above equation,
this is essentially the self-consistent gap equation from the BCS theory.
Because of the sign change, an isotropic $ s $-wave solution is impossible,
since the gap equation is not convergent.
However, since $ \chi( {\bf q } )$ is peaked near $ {\bf Q} = \left( \pi /a, \, \pi/a \right) $,
pairing interaction relates the gap at momenta $ {\bf k } $ and $ { \bf k } + {\bf Q } $.
In this situation, we use the ansatz
$ \Delta ({ \bf k } ) = - \Delta ( {\bf k } + {\bf Q } ) $
therefore eliminating the minus sign.
For tetragonal lattice, this ansatz implies
$ d_{ x^2 - y^2 } $ symmetry of the pairing gap \cite{Annet1992}.

\bibliography{apssamp}

\end{document}